\documentclass[showpacs,aps,superscriptaddress,prb,twocolumn,amsmath]{revtex4}
\usepackage{amssymb}
\usepackage{graphicx,subfigure,float}% Include figure files
\begin{document}

\preprint{APS/123-QED}

%++++++++++++++++++++++++++++++++++++++++++++++++++++
\title{Open Type Nodal line Topological Semimetal in Two Dimensional B$_2$C}
%============================================
\author{P. Zhou}
 \affiliation{Key Laboratory of Low-dimensional Materials and Application Technology, School of Material Sciences and Engineering, Xiangtan University, Xiangtan 411105, China}
 \author{Z. S. Ma}
  \affiliation{Hunan Provincial Key laboratory of Thin Film Materials
and Devices, School of Material Sciences and Engineering, Xiangtan
University, Xiangtan 411105, China}
\author{L. Z. Sun}
 \email{lzsun@xtu.edu.cn}
 \affiliation{Hunan Provincial Key laboratory of Thin Film Materials
and Devices, School of Material Sciences and Engineering, Xiangtan
University, Xiangtan 411105, China}
%============================================
%\keywords{Topological insulator, Bilayer hexagonal lattices, Edge states, Band inversion}
%====================================================================
\date{\today}
%============================================
\begin{abstract}
Topological semimetals, including Dirac semimetals, Weyl semimetals, and nodal line semimetals, receive enormous research interest due to their intrinsic topological nature and fascinating properties. In present work, with the help of density functional theory (DFT), we find that nodal line semimetals and Dirac semimetals can coexist in the low energy electron of two dimensional (2D) monolayer B$_2$C. Intriguingly, besides type-I and type-II Dirac fermions, a kind of open nodal line appears around the Fermi level. The low energy electronic nature of B$_2$C sheet can be described by a tight-binding (TB) model relied on the basis of B-p$_y$, p$_z$ and C-p$_y$, p$_z$ states. One of the most merit of the system is that the energy windows of these four types topological semimetals are different that can be easily distinguished in experiments. Moreover, the system provides an excellent platform for studying the interplay between different type semimetals.\\
\end{abstract}
%\pacs{71.20.-b, 71.70.Ej, 73.20.At}
\maketitle
\begin{figure*}
\includegraphics[trim={0.0in 0.0in 0.0in 0.0in},clip,width=5.0in]{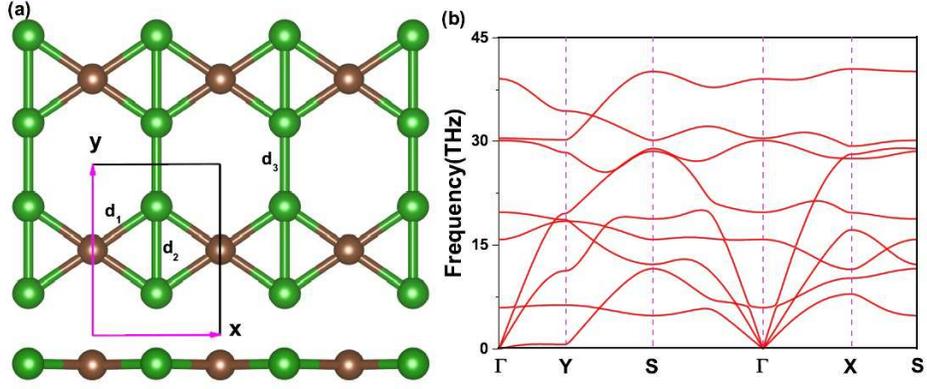}\\
\caption{(color online) (a) The top and side view of monolayer low buckled B$_2$C. (b) Phonon band dispersion of 2D low buckled B$_2$C.}\label{fig1}
\end{figure*}
\begin{figure*}
\includegraphics[trim={0.0in 0.0in 0.0in 0.0in},clip,width=5.0in]{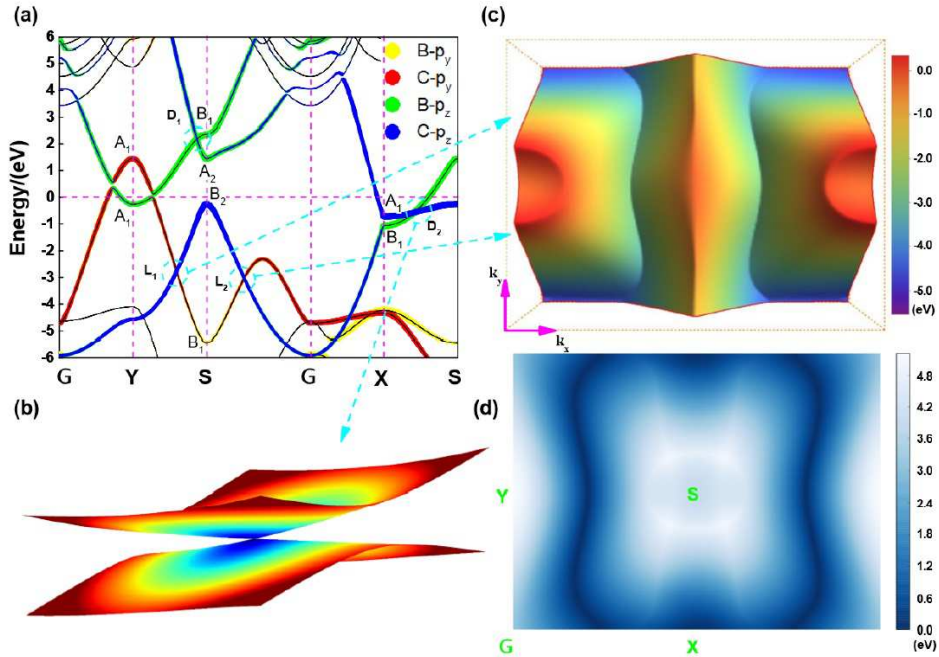}\\
\caption{(color online) Fig.2 (a) Band structure of low buckled B$_2$C. Different colors and spot size represent different orbital contributions and their strength. The point group of all high symmetry point are C$_{2v}$, and irreducible representations (IR) of the bands around Fermi level at high symmetry points(A$_1$, A$_2$, B$_1$ and B$_2$) are included. For any points in high symmetry lines $\Gamma$-Y, Y-S, $\Gamma$-X or X-S(except two high symmetry terminals), the point group symmetry is C$_s$. (b) 3D energy bands of the Dirac cone (type II) along high symmetry line X-S. The color means the absolute value of energy difference between energy of bands and the energy of the Dirac point. (c) 3D energy bands of open type nodal line semimetal around L$_1$ and L$_2$. The direction of reciprocal space are indicated in magenta arrow and the color represent the energy of the three dimensional bands. (d) is the band gap of the open type nodal line of (b) in 2D reciprocal space. The black lines indicate the position of nodal lines in the first Brillouin zone. }\label{fig2}
\end{figure*}
\begin{figure*}
\includegraphics[trim={0.0in 0.0in 0.0in 0.0in},clip,width=5.0in]{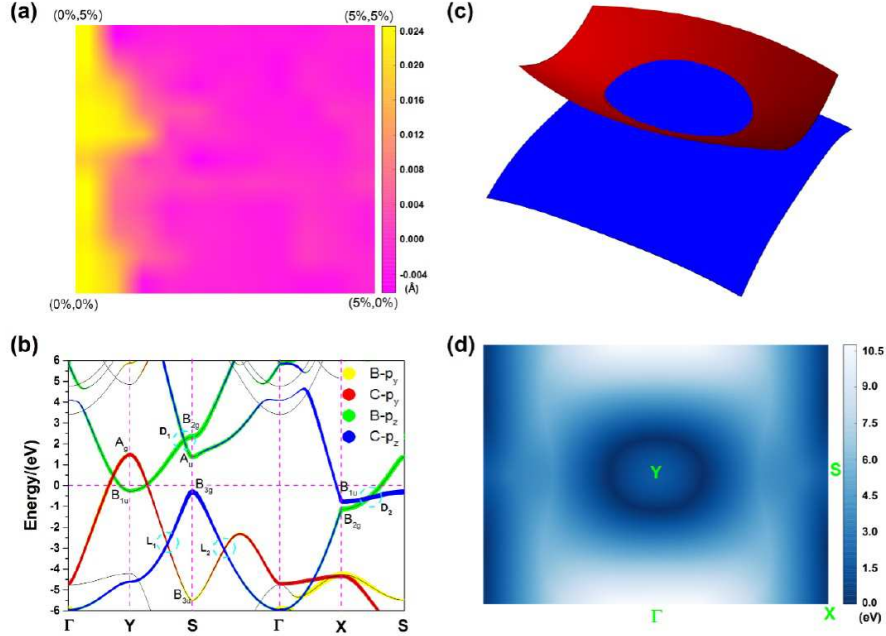}\\
\caption{(color online) (a) The distance $\triangle$h between B and C sublattices along z-direction in function of the stretching strain in-plane for pristine B$_2$C. (b) Band structure of flat B$_2$C. Different colors and spot size represent orbital contributions and their strength. Irreducible representations (IR) of the bands of low energy are included. (c) 3D plot of the closed nodal line around Y point of flat B$_2$C. (d) The band gap of the closed nodal line around Y point corresponding to (c).}\label{fig3}
\end{figure*}
\begin{figure*}
\includegraphics[trim={0.0in 0.0in 0.0in 0.0in},clip,width=6.0in]{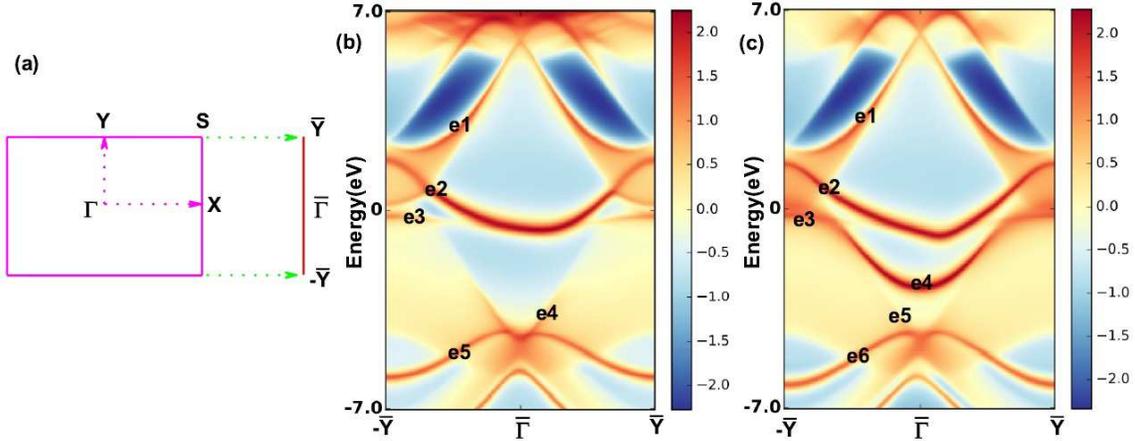}\\
\caption{(color online) (a) 2D Brillouin zone of B$_2$C and its projected 1D Brillouin zone. (b) The surface states of low buckled B$_2$C, we label them as e1-e5 and the correspondence between these edge states and semimetal states are e1$\rightarrow$D$_1$, e3$\rightarrow$D$_2$ and e4$\rightarrow$L$_1$(L$_2$), e2 and e5 are a normal trivial edge states. (c) The surface states of flat B$_2$C, we label them as e1-e6 and the correspondence between these edge states and semimetal states are e1$\rightarrow$D$_1$, e3$\rightarrow$D$_2$ and e5$\rightarrow$L$_1$(L$_2$), e4 is the edge state of closed nodal line in flat B$_2$C and e2 and e6 are a normal trivial edge states. }\label{fig4}
\end{figure*}
\begin{figure*}
\includegraphics[trim={0.0in 0.0in 0.0in 0.0in},clip,width=5.0in]{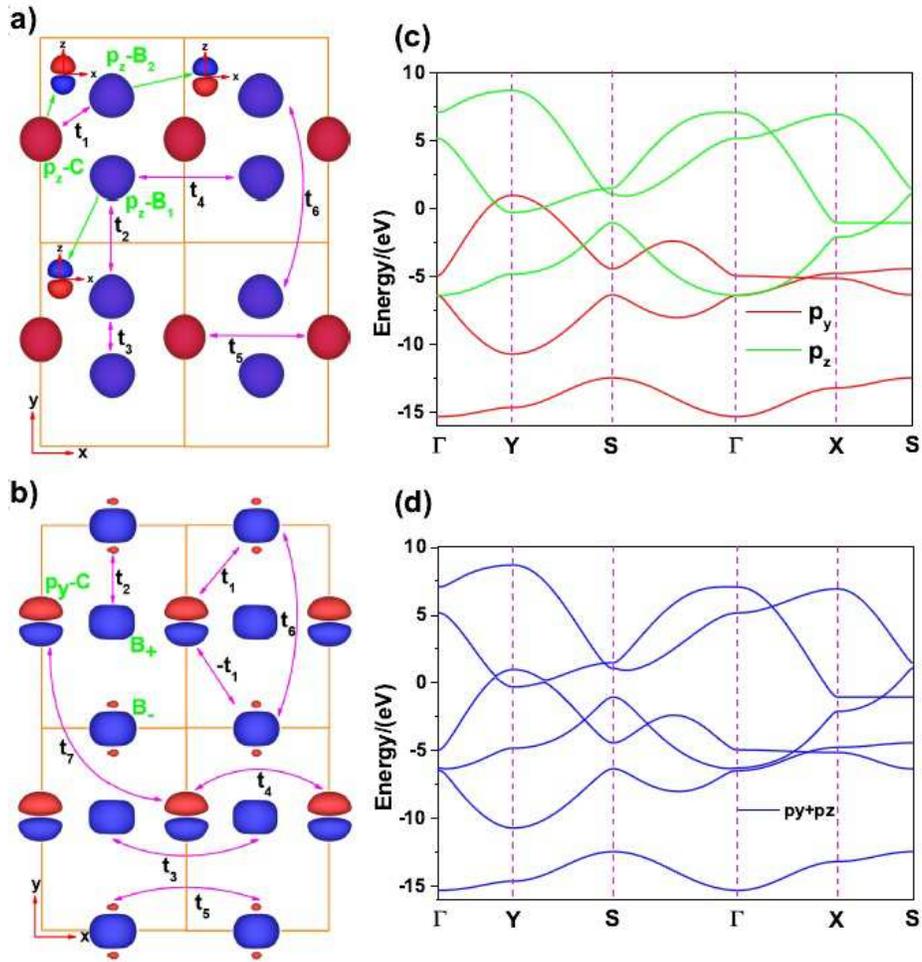}\\
\caption{(color online) (a) The MLWF of p$_z$ states and the scheme of their interaction parameters. (b) The MLWF of p$_y$ states and the scheme of their interaction parameters. (c) The energy bands without the interactions between p$_y$ and p$_z$ states. (d) The energy bands with the interactions between p$_y$ and p$_z$ states.  }\label{fig5}
\end{figure*}
\begin{figure*}
\includegraphics[trim={0.0in 0.0in 0.0in 0.0in},clip,width=6.5in]{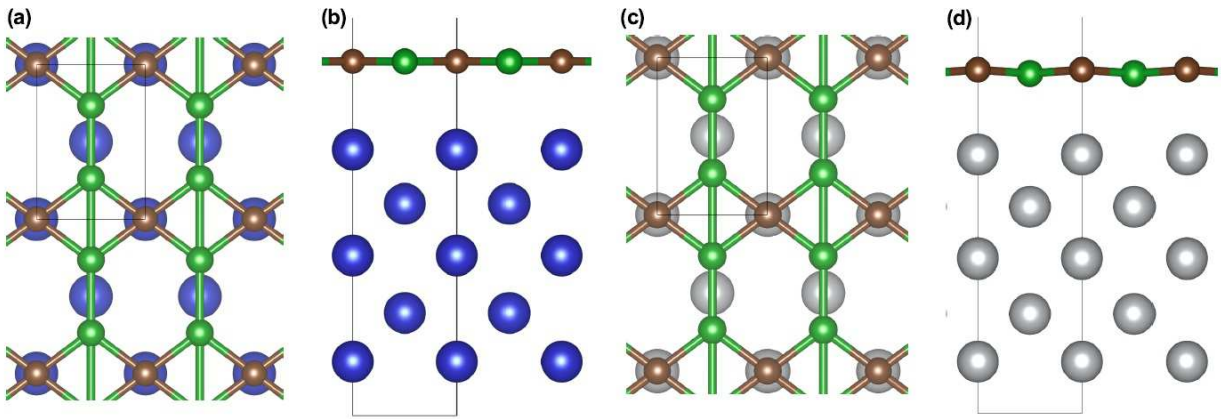}\\
\caption{(color online) The top view (a) and side view (b) of B$_2$C on the surface of Cu(110). The top view (c) and side view (d) of B$_2$C on the surface of Ni(110). }\label{fig6}
\end{figure*}

%++++++++++++++++++++++++++++++++++++++++++++++++++++++++++++
\indent Topological semimetal, following the rising of topological insulator\cite{topormp1,topormp2}, has become the hot topic in condensed matter community. Graphene, as the first experimentally synthesized two dimensional (2D) material, is a typical Dirac semimetal\cite{topodirac} if the weak spin-orbit coupling (SOC) of C atom is neglected. Since the experimental observations of Weyl semimetals\cite{weylexp1,weylexp2,weylexp3}, more and more researchers turn to these special metal states and explore their possible physical phenomena. Some interesting phenomena, including negative magnetoresistance and non-local transport current, have been predicted in this novel type materials\cite{weylthe1,weylthe2}. Following the discovery of type-II Weyl semimetal\cite{t2weyl}, type-II Dirac semimetal was recently proposed\cite{t2dirac1,t2dirac2,t2dirac3}. Its obvious feature is the violating of the Lorentz symmetry comparing with its type-I counterpart. Nodal line semimetal, whose fermi surface is composed by one or several closed lines, also received considerable attention due to its topological nature. The nodal line semimetal was firstly predicated in three dimensional (3D) materials, including  all-carbon Mackay-Terrones crystal\cite{weng1}, anti-perovskite Cu$_3$PdN\cite{Yuprl, Yu2016}, and experimentally observed in 3D materials such as PbTaSe$_2$\cite{3Dnlexp1}, PtSn$_4$\cite{3Dnlexp2}, and ZrSiS\cite{3Dnlexp3}. Lately, several 2D nodal line semimetals, including Honeycomb-Kagome Lattice\cite{2Dnl1}, 2D MX (M=Pd, Pt; X=S, Se, Te)\cite{2Dnl2}, bipartite square lattice\cite{2Dnl3}, and topological (crystalline) insulators thin film\cite{2Dnl4}, have been proposed theoretically. Besides the compatibility with the synthesis of nano-electronics devices, one of the most significant merits of the 2D topological semimetals is that their topological states can be directly characterized by angle-resolved photoemission spectroscopy (ARPES) in experiments, which provides perfect platform for the investigation of the topological nature.\\
%++++++++++++++++++++++++++++++++++++++++++++++++++++++++++++
\indent In present work, with the help of density functional theory (DFT), we find that a novel open type nodal line topological semimetal can coexist with type-I and type-II Dirac semimetal in 2D low buckled B$_2$C. Further investigation also proves that closed type nodal line, open type nodal line, and type-I and type-II Dirac fermions can be found simultaneously in the low energy electronic states of flattened B$_2$C. Moreover, the four types of topological states appear in different energy windows that can be observed and compared in the same experimental observation. With the help of the analysis of the edge states and effective model of the material, the origin of the topological nature was explained. Based on maximally localized Wannier functions, tight-binding model was constructed to investigate the formation of electronic states around Fermi level. Our further particle swarm optimization calculations prove that 2D B$_2$C is feasibly synthesized on the surface of Cu(110) or Ni(110) in experiments, which is crucial premise for the study of the 2D semimetals. \\
%++++++++++++++++++++++++++++++++++++++++++++++++++++++++++++
\indent First-principles calculations were carried out with the Vienna ab initio simulation package (VASP)\cite{vasp1,vasp2}. The GGA of Perdew-Burke-Ernzerhof type exchange-correlation functional was used to simulate electronic interactions. OptB88-vdW exchange-correlation functional\cite{vdw1,vdw2} was applied to consider the Van der Waals interaction for 2D B$_2$C synthesis on the metal substrate. Spin-orbit coupling (SOC) was neglected owing to the light elements of B and C in the material. The cutoff energy for plane-wave expansion was 500 eV and in the self-consistent process $15 \times 11 \times 1$ k-point mesh was used. The crystal structures were fully relaxed until the residual force on each atom was less than 0.01 eV/{\AA}. A vacuum of 15 {\AA} was considered to minimize the interactions between the neighboring periodic images. Phonon spectrum\cite{phonopy} was employed to investigate dynamic stability of the systems. To explore the edge states, we adopt the code of WannierTools\cite{Wu2017}, which cuts the maximally localized Wannier functions (MLWF) Hamiltonian to a semi-infinite form and the MLWFs were generated by using the wannier90 code\cite{wan90}.\\
%++++++++++++++++++++++++++++++++++++++++++++++++++++++++++++
\indent The intrinsic 2D B$_2$C enjoy low buckled configuration between B and C sublattice. The vertical distance between them is 0.025 \AA, which is less than previous reports\cite{B2C_1,B2C_2}. The crystal constants along x and y direction are 2.58 {\AA} and 3.42 {\AA}, respectively. All bond lengths labeled in Fig.\ref{fig1}(a) are d$_1$ = 1.557 \AA, d$_2$ = 1.741 {\AA}, and d$_3$ = 1.683 \AA, respectively. The space group of low buckled B$_2$C is Pmm2(No. 25) with the point group C$_{2v}$, in which only 2-fold rotation along z axis and two mirror symmetries (they are perpendicular to x and y axis) exist. According to the Bader charge analysis\cite{bader}, C atom obtains 0.1 |e| from two equivalent B atoms. The thermodynamic stability have been proved by previous work\cite{B2C_1}. They predicted the melt point of B$_2$C monolayer between 2000 and 2500 K with first-principles molecular dynamic computation. To prove the dynamic stability of low buckled B$_2$C, we calculate the phonon dispersion along high symmetry lines and present the result in Fig.\ref{fig1}(b). The results show that there is not  imaginary frequency in the reciprocal space, therefore, B$_2$C is dynamically stable.\\
%++++++++++++++++++++++++++++++++++++++++++++++++++++++++++++
%++++++++++++++++++++++++++++++++++++++++++++++++++++++++++++
\indent We present the orbit-projected energy band for low buckled B$_2$C in Fig.\ref{fig2}(a). The result clearly reveals that the electronic states of low buckled B$_2$C around Fermi level are mainly decided by the B-p$_{y,z}$ and C-p$_{y,z}$. For the electronic states near the Fermi level around Y point, the electron-like conduction band originates from B-p$_z$ orbitals, whereas the hole-like valence band derives from C-p$_y$. There are small band gaps along high symmetry line due to the interaction of these two orbits (both of them show same irreducible representations (IR) of A$_1$). Moreover, there are several energy band crossings for the low energy electronic states along the high symmetry lines, we labeled them as D$_1$, D$_2$, L$_1$, and L$_2$ in Fig.\ref{fig2}(a). The point group of $\Gamma$ point in reciprocal space for low buckled B$_2$C is C$_{2v}$ with two mirror symmetry planes that is perpendicular to the 2D plane. All the k-point in high symmetry line $\Gamma$-X, $\Gamma$-Y, X-S, and Y-S are share the same point group: C$_s$. As for such point group, there are only two IRs of A$^{'}$ and A$^{''}$, and their mirror operator eigenvalues are opposite. If two energy bands with different IR along the four high symmetry lines as mentioned above, the band cross would unavoidable produces semimetal state. The analysis of the IR of all the four band crossings D$_1$, D$_2$, L$_1$, and L$_2$ indicates that all of them meet the requirement behaving as semimetal. Considering the symmetry of the k-points of the crossings, these four semimetal states are all protected by the mirror symmetry. The 3D energy bands around the band crossing points indicate that D$_1$ is type-I Dirac point and D$_2$ is the latest reported type-II Dirac cone with the broken Lorentz invariance. The 3D energy band of the type-II Dirac cone is shown in Fig.\ref{fig2}(b). Two recent works reported that this topological Dirac (type-I and type-II) points can also be realized in photonic systems under special nonsymmorphic symmetry\cite{photo1,photo2}, which provide an feasible way to adjust this topological semimetal states.\\
%++++++++++++++++++++++++++++++++++++++++++++++++++++++++++++
\indent Intriguingly, the band crossing points of L$_1$ and L$_2$ in the 3D energy bands as shown in Fig.\ref{fig2}(c) and (d) (the position of black lines in Fig.\ref{fig2}(d) represent the zero gap location of the open type nodal line in 2D reciprocal space) indicate that they expand to the edge of the first Brillouin zone and form a novel open type nodal line semimetal, which is recently proposed in three dimensional materials with special space-group symmetry and time-reversal symmetry\cite{open3D}. However, the dimensions of materials and origin of open nodal line are totally different in these two cases. For B$_2$C, it is two dimensional materials and it's open nodal lines mainly come from strong anisotropic in-plane interaction. The IRs of these two bands around the high symmetry point S are B$_{1}$ and B$_{2}$ showing opposite mirror eigenvalues. Moreover, the two bands inverse around the S point between L$_1$ and L$_2$. Such energy band inversion shows strong in-plane anisotropic between x and y directions. The anisotropy produces the energy band crossing points on both sides of line X-$\Gamma$ unconnected each other along k$_x$ orientation and form two separate open type nodal line. The work of Takahashi et al\cite{open3D} concentrate on how nodal line appears under special non-symmorphic symmetry, and they proved that open nodal line may come from exchange glide eigenvalues between two high-symmetry lines. It must be emphasized that the open nodal lines mainly mean energy band crossing lines around high symmetry point do not form closed ring in first Brillouin zone which is independent of the choice of the Brillouin zones. \\
%++++++++++++++++++++++++++++++++++++++++++++++++++++++++++++
%++++++++++++++++++++++++++++++++++++++++++++++++++++++++++++
\indent As mentioned above, the fluctuation between B and C sublattices of low buckled B$_2$C is very small (0.025 \AA), and this z-direction fluctuation can be removed by applying tiny stretching strain. Although, the phonon spectrum as shown in Fig.S1 in the supplementary materials indicates that the isolated flat B$_2$C is dynamically instable, the pure flat B$_2$C can be realized on appropriate substrate or curved substrate. In the following part, we will study the electronic structures of isolated pure flat B$_2$C, and the substrate will be discussed below. The distance $\triangle$h between B and C sublattices along z-direction in function of the stretching strain in x- and y-directions is shown in Fig.\ref{fig3}(a). The results indicate that when the strain along x orientation is larger than 2\%, the $\triangle$h is close to zero (The magenta zone in the figure represents $\triangle$h can be neglected). The results indicate that with stretching strain, the low buckled B$_2$C can be feasibly produced to pure flat one. The space group of flat B$_2$C change to Pmmm (No. 47) with the point group D$_{2h}$. The band structure of pure flat B$_2$C is shown in Fig.\ref{fig3}. In comparison with that of low buckled system, the type-I and type-II Dirac point as well as the open type nodal line around S point in low buckled B$_2$C are all reserved in the flat system. However, the two gaps around Y point of the low buckled B$_2$C near Fermi level are all closed. The 3D energy band around Y point in reciprocal space as shown in Fig.\ref{fig3}(c) indicates that the bands form a nodal loop around Y point. In the figure, the red and blue curved surfaces represent the bands contributed by B-p$_z$ and C-p$_y$, respectively. With the help of group analysis, we find that the point group of four time inversion invariance point in reciprocal space of low buckled B$_2$C is turned to D$_{2h}$ for the flat B$_2$C. In this point group, all three coordinate axes are two-fold rotation axis. As shown in Fig.\ref{fig3}(a), the IR of the two energy bands near Fermi level around Y point are B$_{1u}$ and A$_g$ now, respectively. Under the operator of spatial inversion symmetry, the parity of the wavefunctions of these two bands are opposite, meanwhile, time-reversal symmetry always exists in low buckled or flat B$_2$C. Therefore, the wavefunctions can not interact with each other around the Y producing the gaps close and then forming nodal line around this point. Namely, for the flat B$_2$C, the nodal loop around Y point is protected by the spatial inversion symmetry and time-reversal symmetry at the same time, which prohibit the interaction between p$_z$ and p$_y$ states..\\
%++++++++++++++++++++++++++++++++++++++++++++++++++++++++++++
%++++++++++++++++++++++++++++++++++++++++++++++++++++++++++++
\indent Here we will firstly discuss the origin of the topological properties of Dirac semimetal states of D$_1$ and D$_2$. Under the time-reversal symmetry ($\mathcal{T}$) and spatial inversion symmetry($\mathcal{I}$), the Zak phase is quantized and topological analysis is meaningful\cite{topo2D}, therefore, we concentrate on the case of flat B$_2$C. According to previous work\cite{topo2D}, 2D semimetal states can be divided into two categories: trivial and non-trivial Z$_2$ invariant semimetal. They are decided by the parity of occupied Bloch wave function at $\mathcal{T}$ points. The Dirac semimetal states D$_1$ and D$_2$ of 2D B$_2$C are formed by the crossing of two bands with different parity as shown in Fig.\ref{fig3}(b), their topological properties should be contributed by the parity of the two bands in reciprocal space. In the unit cell of B$_2$C, two B atoms and one C atom contribute totally ten electrons. Ignoring the freedom of spin, ten valence electrons will occupy five lowest energy bands. For the two crossing bands of D$_1$ and D$_2$, we assume the band whose energy is higher than the other belongs to the conduction band. According to the work of Miert et al\cite{topo2D},  if the Dirac point locate in the high symmetry lines with two time-reversal invariant terminals, the quantized Zak phase can be simplified as the parity of the full Bloch wave function at the time-reversal invariant points. The concrete formula is:
%+++++++++++++++++++++
\begin{gather}
\chi_1 = \prod_{i \in occ}\xi_i(\vec{b_1})\xi_i(\vec{b_2})\label{eq1}
\end{gather}
%+++++++++++++++++++++
where $\xi_i$ is the occupied energy band, $\vec{b_1}$ and $\vec{b_2}$ represent the beginning and the ending point of any high symmetry line, respectively, for with one Dirac point on it. For D$_2$, the high symmetry line is X-S, then the $\vec{b_1}$ and $\vec{b_2}$ are X and S points, respectively. However, for the case of D$_1$, the high symmetry line is Y-S, and the $\vec{b_1}$ and $\vec{b_2}$ indicate Y and S high symmetry points, respectively. The result of Eq.\ref{eq1} is -1(1) indicating the system topological non-trivial(trivial). As for D$_2$, the parities of valence bands at X and S high symmetry point are (+1, -1, -1, -1, +1) and (+1, -1, +1, -1, +1), respectively. The parities of valence bands at Y and S for D$_1$ are (-1, +1, +1, +1, -1, +1) and (+1, -1, +1, -1, +1, -1), respectively. Based on Eq.\ref{eq1}, both D$_1$ and D$_2$ semimetal states are topological non-trivial, which also are proved by our further Berry phase calculation around these two points. If we adapt the anticlockwise circular k-path in reciprocal space, the Berry phase of D$_1$ and D$_2$ are $\pi$ and -$\pi$, respectively. To further prove the topological nature of these semimetals states, we calculated the edge states with the Green's function method and semi-infinite MLWF Hamiltonian. The results for low buckled and flat B$_2$C are shown in Fig.\ref{fig4} (b) and (c), respectively. All edge states around Fermi level are labeled as e1 to e6 in Fig.\ref{fig4}(c). By comparing the number of edge states between Fig.\ref{fig4}(b) and Fig.\ref{fig4}(c), e4 is the edge state of closed nodal line in flat B$_2$C. According the energy range of other semimetal states as shown in Fig.\ref{fig2}(a) and Fig.\ref{fig3}(b), we can sum up the correspondence between these edge states and semimetal states as: e1 $\rightarrow$ D$_1$, e3 $\rightarrow$ D$_2$, e5 $\rightarrow$ L$_1$(L$_2$). However, after analyzing the edge states in detail with different symmetry, we found the e2 and e6 are a normal trivial edge states derived from the symmetric broken of the ribbon edge.\\
%++++++++++++++++++++++++++++++++++++++++++++++++++++++++++++
%++++++++++++++++++++++++++++++++++++++++++++++++++++++++++++
\indent In this part, we derive the effective model to two different type nodal line of flat B$_2$C. Generally, a two-band model can be written as following form\cite{Yu2016}:
%++++++++++++++++++++++
\begin{gather}
H(\textbf{k}) = g_0(\textbf{k})\sigma_0 + g_1(\textbf{k})\sigma_x + g_2(\textbf{k})\sigma_y + g_3(\textbf{k})\sigma_z
\end{gather}
%++++++++++++++++++++++
where $\sigma_0$ is 2 $\times$ 2 unit matrix, and $\sigma_{x,y,z}$ are Pauli Matrices. $g_{0,1,2,3}$($\textbf{k}$) denote real function of $\textbf{k}$(k$_x$,k$_y$). Under space-inversion symmetry operator $\emph{\^{P}}$ and time-reversal symmetry operator $\emph{\^{T}}$, the effective Hamiltonian must satisfy
%++++++++++++++++++++++
\begin{gather}
\emph{\^{P}}H(\textbf{k})\emph{\^{P}}^{-1} = H(-\textbf{k}), \ \   \emph{\^{T}}H(\textbf{k})\emph{\^{T}}^{-1} = H(-\textbf{k})
\end{gather}
%++++++++++++++++++++++
Because the two bands we considered own opposite parity if omit the freedom of spin, hence the operator $\emph{\^{P}}$ and $\emph{\^{T}}$ can be chosen as $\sigma_z$ and $\emph{\textbf{K}}$. $\emph{\textbf{K}}$ represents the complex-conjugate operator. After substituting them into above formula, we can obtain that $g_1(\textbf{k})$ = 0, $g_2(\textbf{k})$ is an odd function of $\textbf{k}$  and $g_{0,3}(\textbf{k})$ are even functions of $\textbf{k}$. Moreover, two additional mirror symmetries would require the Hamiltonian satisfying: k$_x$ $\leftrightarrow$ -k$_x$; k$_y$ $\leftrightarrow$ -k$_y$. If we only expand the Hamiltonian to two order of $\textbf{k}$, we have
%++++++++++++++++++++++
\begin{align}
& g_0(\textbf{k}) = a_0 + a_1{k_x}^2 + a_2{k_y}^2\\
& g_2(\textbf{k}) = 0\\
& g_3(\textbf{k}) = m_0 + m_1{k_x}^2 + m_2{k_y}^2
\end{align}
%++++++++++++++++++++++
Because the $g_0(\textbf{k})$ only breaks the particle-hole symmetry and affects the energy dispersion of the nodal line, we only need to consider the case of $g_3(\textbf{k})$ = 0. The parameters $m_0$, $m_1$, and $m_2$ of $g_3(\textbf{k})$ are independent each other due to no additional symmetry restriction. When $m_0$ and $m_{1,2}$ have opposite signs, the equation is equivalent to ellipse equation ($\frac{{k_x}^2} {\frac{-m_0}{m_1}} + \frac{{k_y}^2}{\frac{-m_0}{m_2}} = 1$). If the length of the two principal axes ($\sqrt{\frac{-m_0}{m_1}}$ and $\sqrt{\frac{-m_0}{m_2}}$ ) of the ellipse are both shorter than corresponding basis vector of 2D reciprocal space, a nodal loop will form. The closed nodal line around Y point for flat B$_2$C is the case. If the length of any principal axis of the ellipse is longer than corresponding primitive vector, two non-closed nodal lines would be formed, this is the case of open type nodal lines in both low buckled and flat B$_2$C. It is worth noting that there is a third case that the length of all principal axis of the ellipse are longer than corresponding primitive vector and one of them is shorter than the diagonal of the 2D rectangle Brillouin zone. In such case, four separated nodal lines will form in the first Brillouin Zone, which is very similar with the 2D boundary plane of Cu$_3$NPd\cite{3Dnlthe1}, even though it is 3D materials. Below, we will use TB Hamiltonian to explain the formation of the two types nodal lines in B$_2$C.\\
%++++++++++++++++++++++++++++++++++++++++++++++++++++++++++++
\indent According to the atom-projected energy bands around Fermi level as shown in Fig.\ref{fig2}(a) and Fig.\ref{fig3}(b), we know that the low energy electronic states mainly come from p$_y$ and p$_z$ states of B and C atoms. To understand the formation of semimetal states in 2D B$_2$C, we can construct a six-band TB model with the help of MLWFs. The hamiltonian can be written as:
%+++++++++++++++++
\begin{gather*}
H = \begin{pmatrix}H_{1} & {\triangle}H \\ {\triangle}H^{*} & H_{2}  \end{pmatrix}
\end{gather*}
%+++++++++++++++++
where $H_1$ and $H_2$ are constructed with the basis of p$_y$ and p$_z$ states of all atoms, respectively. ${\bigtriangleup}H$ is the perturbation to describe the interaction between p$_y$ and p$_z$ states. For p$_z$ states, the center of Wannier function locates near the position of atoms, therefore, the basis of them can be set as traditional TB model ({p$_z$}-C, p$_z$-B$_1$, and p$_z$-B$_2$) as shown in Fig.\ref{fig5}(a). However, the p$_y$ states of B atoms interact with each other and form bonding $\frac{1}{\sqrt{2}}$(p$_y$-B$_1$ + p$_y$-B$_2$)$\rightarrow$ B$_{+}$ and antibonding $\frac{1}{\sqrt{2}}$(p$_y$-B$_1$ - p$_y$-B$_2$)$\rightarrow$ B$_{-}$ due to the mirror symmetry. The bonding state center locates at the midpoint between B$_1$ and B$_2$ atoms, whereas the antibonding state center is at the midpoint of primitive vector along x-axis as shown in Fig.\ref{fig5}(b). Therefore, the basis can be set as (p$_y$-C,B$_{+}$, and B$_{-}$). With the two basis, the $H_1$ and $H_2$ in TB model can be formulated as follows:
%+++++++++++++++++
\begin{gather}
H_n = \sum_{i}\epsilon_{in}c_{in}^{\dag}c_{in}+\sum_{i{\neq}j}t_{{\alpha}n}c_{in}^{\dag}c_{jn}
\end{gather}
%+++++++++++++++++
where n = 1 (2) represent the three bands model with the basis of p$_y$ (p$_z$) orbits, $\epsilon_{in}$ is on-site energy, $c_{in}^{\dag}$ ($c_{jn}$) represents the creation (annihilation) operator of electrons at site i (j). $t_{{\alpha}n}$($\alpha$ = 1-7 for p$_y$ model and $\alpha$ = 1-6 for p$_z$ model) are the hopping parameter between the i-th and j-th sites. The details of the parameters used in our present work can be found in the Tab.S1 in the supplementary materials. When ${\bigtriangleup}H$ is 3$\times$ 3 zero matrix, namely the interactions between p$_y$ and p$_z$ orbits are omitted, we plot the band structure of $H$ in Fig.\ref{fig5}(c). The red and green lines are, respectively, come from the eigenvalues of $H_1$ and $H_2$ and they are well match with the first-principles energy band of Fig.\ref{fig3}(b). When we add a weak interaction between p$_y$ and p$_z$ orbits, band gap will open for the nodal line around Y high symmetry point. Here we take the interaction parameter of -0.1 between B$_{-}$ and its nearest neighbor p$_z$-B$_1$ or p$_z$-B$_2$ as an example, and the band structures along high symmetry lines are shown in Fig.\ref{fig5}(d). Clearly, local band gaps open for the closed nodal line around Y. With the help of energy band structures derived from TB model, we can draw the following conclusions: The two types of nodal lines in flat B$_2$C come from the energy band crossing between p$_y$ and p$_z$ orbits, but the two Dirac semimetal states are only derived from p$_z$ orbitals. The disappearance of the interaction between p$_y$ and p$_z$ orbits for flat B$_2$C can well describe the existence of the closed nodal line around Y point. Moreover, whether or not ${\bigtriangleup}H$ matrix exists, the interaction between these two types of orbitals would not break the mirror symmetry along x and y axis. Therefore, the open type nodal line and two Dirac semimetal states are all determined by the mirror symmetries and independent of  ${\bigtriangleup}H$.\\
%++++++++++++++++++++++++++++++++++++++++++++++++++++++++++++
%++++++++++++++++++++++++++++++++++++++++++++++++++++++++++++
%++++++++++++++++++++++++++++++++++++++++++++++++++++++++++++
%++++++++++++++++++++++++++++++++++++++++++++++++++++++++++++
\indent Now, we will discuss the proper substrate to synthesize the monolayer B$_2$C. It is well known that crystal constant matching between substrates and 2D materials is an important prerequisite for synthesizing monolayer 2D materials. By comparing the crystal constants between different metal surface and B$_2$C, we find that the crystal constants of B$_2$C match well with Cu(110) surface (the mismatches are -0.95\% and 5.2\% in x and y orientations, respectively) and Ni(110) surface (the mismatches are -3.5\% and 2.8\% in x and y orientations, respectively, where the negative percentage mean the lattice constant of substrate is shorter than corresponding lattice constant of low buckled B$_2$C.). To obtain the most stable B$_2$C structure by self-assembled process on these two metal surfaces, we adapt the recently developed surface reconstruction prediction method\cite{calypso1} based on particle swarm optimization (PSO)\cite{calypso2}, which has been recently utilized to simulate the growth of monolayer borophene on different metal substrate\cite{B_acie}. The most stable B$_2$C on the surface of Cu(110) and Ni(110) are present in the Fig.\ref{fig6}. The results indicate that the fundamental structure of 2D B$_2$C is well reserved in comparison with free standing monolayer film. The buckle between B and C sublattice of B$_2$C is negligible when it on the surface of Cu(110) realizing the flat B$_2$C. However, when it on the surface of Ni(110), the buckle reaches 0.127 {\AA} obtaining the buckled B$_2$C. We also present other typical adsorption structures on the surface of Cu(110) with higher energy in Fig.S2 in the supplementary materials. Although the results indicate that the B or C atoms on the grooves of Cu(110) or Ni(110) surface are not energy favorable, the energy difference between different adsorption configurations are within several meV. The above results indicate that the Cu(110) or Ni(110) surfaces are promising substrate for growing the monolayer B$_2$C. However, the p$_z$ states of B and C atoms are largely affected due to the interaction between B$_2$C and metal substrate, by which the topological properties of the systems can not be reserved. To guarantee the topological metal states, we construct a van der Waals bilayer B$_2$C on the surface of the substrate, the detailed electronic structure analysis are presented in Fig.S3 in the supplementary materials. The projected energy bands of the superficial B$_2$C indicate that the topological metal states are well reserved.\\
%++++++++++++++++++++++++++++++++++++++++++++++++++++++++++++
%++++++++++++++++++++++++++++++++++++++++++++++++++++++++++++
\indent The most interesting nature of B$_2$C is that its different type semimetal states around Fermi level locate different energy windows, that can be easily distinguished in experiments. Nowadays, the nodal line semimetal state was only indirectly observed in 3D materials in experiments with angle-resolved photoemission spectroscopy (ARPES), which is largely affected by the orientation and termination of the surface. Therefore, we expect the 2D B$_2$C would be the choice example to directly observe the state of nodal line semimetal. Moreover, for low buckled B$_2$C, there are three types semimetal states coexistence (including type-I and type-II Dirac semimetal and open type nodal line semimetals), whereas for flat B$_2$C, there are four types semimetal states simultaneous existence (including type-I and type-II Dirac semimetal and open and close type nodal line semimetals). All the coexisting semimetal states located at different momenta might couple to each other through electron interactions, namely, the B$_2$C system will provide an excellent platform for studying the interplay between different types semimetals.\\
%++++++++++++++++++++++++++++++++++++++++++++++++++++++++++++
%\section*{CONCLUSION}
\indent In summary, we have found four semimetal states coexist in the low energy electronic state of 2D B$_2$C with negligible SOC. These semimetal states are protected by $\mathcal{T}$, $\mathcal{I}$, or mirror symmetry. It is worth mentioning that the open type nodal line is firstly proposed in 2D materials, and it is rooted in the strong in-plane anisotropy interaction. Two band effective model and six-orbital TB model have been constructed to describe the details of anisotropy interaction and the origin of topological semimetal states. The different types semimetal states of 2D B$_2$C around Fermi level locate different energy window that can be easily distinguished in experiments. Moreover, the system provides an excellent platform for studying the interplay between different types semimetals.\\
%============================================================
\begin{acknowledgments}
This work is supported by the National Natural Science Foundation of China (Grant No. 11574260) and scientific research innovation project of Hunan Province(CX2015B219).
\end{acknowledgments}

%\bibliography{Zhou}
%\section{Table of Contents}
%\begin{figure}
%  \includegraphics[width=4.5in]{TOC.eps}\\
%\end{figure}
%\section{Table of Contents}
%\begin{figure}
%  \includegraphics[width=4.5in]{TOC.eps}\\
%\end{figure}
%\end{document}

%\begin{references}{1}


\end{document}